\documentclass[conference, letterpaper]{IEEEtran}
\IEEEoverridecommandlockouts
\usepackage{cite}
\usepackage{amsmath,amssymb,amsfonts}
\usepackage{algorithmic}
\usepackage{graphicx}
\usepackage{textcomp}
\usepackage{xcolor}
\usepackage[utf8]{inputenc}
\def\BibTeX{{\rm B\kern-.05em{\sc i\kern-.025em b}\kern-.08em
    T\kern-.1667em\lower.7ex\hbox{E}\kern-.125emX}}

\usepackage{tikz}
\usetikzlibrary{decorations}
\usepackage{pgfplots}

\usepackage{rotating}
\usepackage{array}

\usepackage{etoolbox}

\usepackage{tabularx}

\usepackage{url}

\newcolumntype{M}{>{\centering\arraybackslash}m{2cm}}

\newcommand{\vect}[1]{\mathbf{#1}}
\newcommand{\scalar}[2]{\left\langle#1,#2\right\rangle}
\newcommand{\norm}[1]{\left\lVert#1\right\rVert}
\newcommand{\Zqn}[0]{\mathbb{Z}_q^n}

\newtheorem{defn}{Definition}
\newtheorem{thm}{Theorem}

\begin{document}

\newtoggle{color}
\toggletrue{color} 
  
\newtoggle{full}
\toggletrue{full} 

\title{The Asymptotic Complexity of Coded-BKW with Sieving Using Increasing Reduction Factors
\thanks{This work was supported by the Swedish Research Council (Grant No. 2015-04528).}
}

\author{\IEEEauthorblockN{Erik Mårtensson}
\IEEEauthorblockA{\textit{Dept. of Electrical and Information Technology} \\
\textit{Lund University}, Lund, Sweden\\
Email: erik.martensson@eit.lth.se}
}

\maketitle

\begin{abstract}
The Learning with Errors problem (LWE) is one of the main candidates for post-quantum cryptography. At Asiacrypt 2017, coded-BKW with sieving, an algorithm combining the Blum-Kalai-Wasserman algorithm (BKW) with lattice sieving techniques, was proposed. In this paper, we improve that algorithm by using different reduction factors in different steps of the sieving part of the algorithm. In the Regev setting, where $q = n^2$ and $\sigma = n^{1.5}/(\sqrt{2\pi}\log_2^2 n)$, the asymptotic complexity is $2^{0.8917n}$, improving the previously best complexity of $2^{{0.8927n}}$. When a quantum computer is assumed or the number of samples is limited, we get a similar level of improvement.
\end{abstract}

\iftoggle{full}{%
	
}{
	\textit{A full version of this paper is accessible at: }
	\url{https://arxiv.org/abs/1901.06558}
}

\section{Introduction}
Given access to large-scale quantum computers, Shor's algorithm solves both the integer factoring problem and the discrete logarithm problem in polynomial time. To remedy this, National Institute of Standards and Technology (NIST) has an ongoing competition to develop post-quantum cryptosystems \cite{NIST}. One of the main underlying mathematical problems in the competition is the Learning with Errors problem (LWE).

The LWE problem was introduced by Regev in \cite{STOC:Regev05}. It has some really nice features, such as a reduction from average-case LWE instances to worst-case instances of hard lattice problems. An application of LWE is Fully Homomorphic Encryption (FHE) \cite{STOC:Gentry09}. An important special case of LWE is the Learning Parity with Noise problem (LPN), essentially a binary version of LWE with Bernoulli distributed noise. 

There are mainly three types of algorithms for solving the LWE problem. For surveys on the concrete and asymptotic complexity of these algorithms see \cite{ConcreteLWE} and \cite{AsymptoticLWE} respectively.

The first type is the Arora-Ge algorithm, which was introduced in \cite{AroraGe}, and then improved in \cite{EPRINT:ACFP14}. This type of algorithm is mostly applicable when the noise is too small for Regev's reduction proof to apply \cite{STOC:Regev05}.

The second type of approach is lattice-based algorithms, where LWE is transformed into a lattice problem and then solved by methods like lattice-reduction, lattice sieving and enumeration.  Lattice-based algorithms are currently the fastest algorithms in practice, and have the advantage of not needing an exponential amount of samples. For more details see \cite{ConcreteLWE} and the references therein. 

The third type of approach is the Blum-Kalai-Wasserman (BKW) set of algorithms. These will be the focus of this paper.

The BKW algorithm was introduced in \cite{STOC:BluKalWas00} as the first sub-exponential algorithm for solving the LPN problem. It was first used to solve the LWE problem in \cite{Albrecht2015}. This was improved in \cite{PKC:AFFP14} using Lazy Modulus Switching (LMS). Further improvements were made in \cite{C:GuoJohSta15, C:KirFou15} by using a varying step size and a varying degree of reduction. In \cite{AC:GJMS17} coded-BKW with sieving was introduced, where lattice sieving techniques were used to improve the BKW algorithm. The full version in \cite{FullCodedBKWWithSieving} improved the coded-BKW with sieving algorithm by finding the optimal reduction factor used for lattice sieving. 

In this paper we further improve upon the coded-BKW with sieving algorithm by increasing the reduction factor for each step of the algorithm. We achieve a record low time complexity of $2^{0.8917n}$ in the Regev setting; that is, when $q = n^2$ and $\sigma = n^{1.5}/(\sqrt{2\pi}\log_2^2 n)$. The previous best result was $2^{0.8927n}$ from \cite{FullCodedBKWWithSieving}. Also if a quantum computer is assumed or the number of samples is limited we get a similar level of improvement. 

The remaining parts of the paper are organized the following way. We start off in Section \ref{sec:Preliminaries} by introducing the LWE problem. In Section \ref{sec:BKW} we go over the previous versions of the BKW algorithm, when used for solving the LWE problem. In Section \ref{sec:NewAlgorithm} we introduce the new algorithm and in Section \ref{sec:Asymptotics} we cover the asymptotic complexity of it and other algorithms for solving LWE. We show our results in Section \ref{sec:Results} and conclude the paper in Section \ref{sec:Conclusions}.

\section{Preliminaries}
\label{sec:Preliminaries}

Let us define the LWE problem.

\begin{defn}[LWE]

Let $n$ be a positive integer, $q$ a prime. Let $\bf s$ be a uniformly random secret vector in $\Zqn$. Assume access to $m$ noisy scalar products between $\bf s$ and known vectors $\bf a_i$, i.e.
\begin{equation*}
z_i = \scalar{{\bf a_i}}{{\bf s}} + {e_i},
\end{equation*}
for $i = 1, \ldots, m$. The small error terms ${e_i}$ are discrete Gaussian distributed with mean 0 and standard deviation $\sigma$. The (search) LWE problem is to find the secret vector $\bf s$.

\end{defn}

In other words, when solving LWE you have access to a large set of pairs $(\mathbf{a_i}, z_i)$ and want to find the corresponding secret vector $\mathbf{s}$. In some versions there are restrictions on the number of samples you have access to.

\section{BKW}
\label{sec:BKW}

BKW was introduced as the first sub-exponential algorithm for solving LPN (essentially LWE with $q = 2$) in \cite{STOC:BluKalWas00}. It was first used for solving LWE in \cite{Albrecht2015}.

\subsection{Plain BKW}

The BKW algorithm consists of two steps, dimension reduction and guessing. 

\subsubsection{Reduction}
Map all the samples into categories, such that the first $b$ positions get canceled when adding/subtracting a pair of $\mathbf{a}$ vectors within the same category.

Given two samples $([\pm \mathbf{a_0}, \mathbf{a_1}], z_1)$ and $([\pm \mathbf{a_0}, \mathbf{a_2}], z_2)$ within the same category. By adding/subtracting the $\mathbf{a}$ vectors we get

$$\mathbf{a_{1, 2}} = [\underbrace{\begin{matrix} 
	0&0&\cdots&0
\end{matrix}}_{b \textnormal{ symbols}}\begin{matrix}
&*&*&\cdots&*
\end{matrix}].$$

By also calculating the corresponding $z$ value we get $z_{1, 2} = z_1 \pm z_2$. Now we have a new sample $(\mathbf{a_{1, 2}}, z_{1, 2})$. The corresponding noise variable is $e_{1, 2} = e_1 \pm e_2$. Thus the variance of the new noise is $2\sigma^2$, where $\sigma^2$ is the variance of the originial noise. By going through all categories and calculating a suitable amount of new samples we have reduced the dimensionality of the problem by $b$, at the cost of increasing the noise. If we repeat the reduction process $t_0$ times we end up with a dimensionality of $n - t_0b$, and a noise variance of $2^{t_0} \cdot \sigma^2$.

\subsubsection{Guessing}
The final positions of the secret vector $\mathbf{s}$ can be guessed and then each guess can be tested using a distinguisher. The guessing procedure does not affect the asymptotics, but is important for concrete complexity. The guessing procedure was improved in \cite{EC:DucTraVau15} using the Fast Fourier Transform (FFT).

\subsection{Lazy Modulus Switching}
The basic BKW algorithm was improved in \cite{PKC:AFFP14} by Albrecht et al. The main idea there was to map samples that, almost but not completely, canceled each other, into the same category. This technique is called Lazy Modulus Switching (LMS).

By doing this an extra error term gets added in each step. The variance of this noise also doubles in each new reduction step. However, LMS allows us to use a larger step size, allowing us to solve larger LWE problems.

One problem with this version of the algorithm is that the extra added noise of the earlier steps grows in size much more than the noise of the later steps, leading to an uneven noise distribution among the positions of the final samples used for the guessing procedure.

\subsection{Coded-BKW}
The problem with the uneven noise distribution was adressed independently in \cite{C:GuoJohSta15, C:KirFou15}. The idea was to use a small step size and almost reduce the positions in the $\mathbf{a}$ vectors to 0 in the first step, and then gradually increase the step size $n_i$ and use less strict reduction for each step.

In \cite{C:GuoJohSta15} different $q$-ary linear codes $\mathcal{C}_i$ with parameters $[n_i, b]$ were used, to vary the strictness of reduction. That version of BKW is called coded-BKW. For simplicity, consider the first reduction step. Pick two samples, such that the first $n_1$ positions of the $\mathbf{a}$ vectors map to the same codeword $\mathbf{c_0}$ in $\mathcal{C}_1$. In other words, we can write

$$
\begin{array}{ccl}
z_1 &=& \scalar{ [\vect c_0 +  \hat{\vect e}_1, \vect a_{1}]}{\textbf{ s}} + { e_1} \\
z_2 &=& \scalar{[\vect c_0 +  \hat{\vect e}_2,\vect a_{2}]}{\textbf{ s}} + { e_2},
\end{array}
$$

where $\hat{\vect e}_1$ and $\hat{\vect e}_2$ have small Euclidean norms. We can get a new sample by calculating

$$
z_1 - z_2 = \scalar{[\hat{\vect e}_1 - \hat{\vect e}_2, \vect a_1-\vect a_{2}]}{\textbf{ s}} + { e_1} - { e_2}.
$$

Just like when using LMS, using this version of BKW adds an extra noise term, but allows us to use larger step sizes.

\subsection{Coded-BKW with Sieving}
In \cite{AC:GJMS17} an idea for combining BKW with lattice sieving techniques was introduced. Just like in \cite{C:GuoJohSta15, C:KirFou15} in step $i$, samples were mapped into categories based on the current $n_i$ positions. Let $N_i = \sum_{j = 1}^i n_j$. The new idea was to only add/subtract samples within a category such that also the previous $N_{i - 1}$ positions of the resulting $\mathbf{a}$ vector were equally small. This could have been done by looking at all possible pairs and picking only the ones with the smallest values in these positions. However, a more efficient way of doing this was to use lattice sieving techniques to find close pairs of vectors within a category faster.

A micro picture of coded-BKW with sieving can be found in Figure \ref{fig:micro}. After step $i$, the average magnitude of the first $N_i$ positions in the $\mathbf{a}$ vector is less than a constant $B$.

\iftoggle{color}{%
	\definecolor{myred}{rgb}{1, 0.25, 0.21}
\definecolor{mygreen}{rgb}{0.18, 0.8, 0.25}
\definecolor{myblue}{rgb}{0, 0.45, 0.85}
\definecolor{myyellow}{rgb}{1, 0.8, 0}

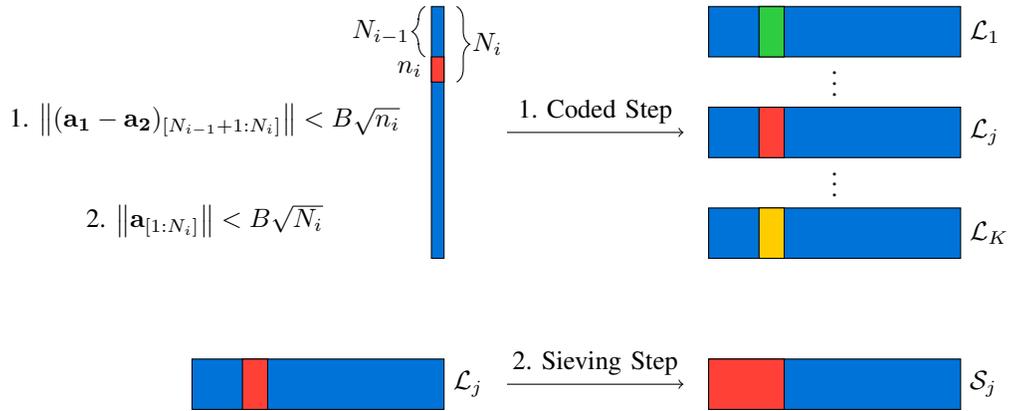
\begin{figure*}[t]
\begin{center}
\begin{tikzpicture}[scale = 0.335]

\draw[black, fill = myblue] (5, 3) rectangle (5.5, 13);
\draw[black, fill = myred] (5, 10) rectangle (5.5, 11);
\node[left] at (5, 10.5) {$n_i$};
\draw [decorate , decoration = {brace, amplitude = 5pt}] (4.75, 11) -- (4.75, 13) {};
\node[left] at (4.5, 12) {$N_{i-1}$};
\draw [decorate , decoration = {brace, amplitude = 5pt}] (6, 13) -- (6, 10) {};
\node[right] at (6.25, 11.5) {$N_i$};

\draw [->] (8, 8) -- (15, 8);
\node[above] at (11.5, 8) {1. Coded Step};

\draw[black, fill = myblue] (16, 11) rectangle (26, 13);
\draw[black, fill = mygreen] (18, 11) rectangle (19, 13);
\node[right] at (26, 12) {${\cal L}_1$};

\node at (21, 10.25) {$\vdots$};

\draw[black, fill = myblue] (16, 7) rectangle (26, 9);
\draw[black, fill = myred] (18, 7) rectangle (19, 9);
\node[right] at (26, 8) {${\cal L}_j$};

\node at (21, 6.25) {$\vdots$};

\draw[black, fill = myblue] (16, 3) rectangle (26, 5);
\draw[black, fill = myyellow] (18, 3) rectangle (19, 5);
\node[right] at (26, 4) {${\cal L}_K$};

\draw[black, fill = myblue] (-4.5, -3) rectangle (5.5, -1);
\draw[black, fill = myred] (-2.5, -3) rectangle (-1.5, -1);
\node[right] at (5.5, -2) {${\cal L}_j$};
\draw [->] (8, -2) -- (15, -2);
\node[above] at (11.5, -2) {2. Sieving Step};

\draw[black, fill = myblue] (16, -3) rectangle (26, -1);
\draw[black, fill = myred] (16, -3) rectangle (19, -1);
\node[right] at (26, -2) {${\cal S}_j$};

\node at (-4, 8.5) {1. $\norm{(\mathbf{a_1} - \mathbf{a_2})_{[N_{i-1}+1:N_i]}} < B\sqrt{n_i}$};

\node at (-4, 4.5) {2. $\norm{\mathbf{a}_{[1:N_i]}} < B\sqrt{N_i}$};

\end{tikzpicture}
\end{center}
\caption{A micro picture of how one step of coded-BKW with sieving works. Slightly changed version of Figure 1 from \cite{FullCodedBKWWithSieving}. Each sample gets mapped to one list ${\cal L}_j$ out of $K$ lists. Sieving is then applied to each list to form new lists ${\cal S}_j$.}
\label{fig:micro}
\end{figure*}

}{
	\begin{figure*}
\begin{center}
\begin{tikzpicture}[scale = 0.335]

\draw[black, fill = white!20!black] (5, 3) rectangle (5.5, 13);
\draw[black, fill = white!80!black] (5, 10) rectangle (5.5, 11);
\node[left] at (5, 10.5) {$n_i$};
\draw [decorate , decoration = {brace, amplitude = 5pt}] (4.75, 11) -- (4.75, 13) {};
\node[left] at (4.5, 12) {$N_{i-1}$};
\draw [decorate , decoration = {brace, amplitude = 5pt}] (6, 13) -- (6, 10) {};
\node[right] at (6.25, 11.5) {$N_i$};

\draw [->] (8, 8) -- (15, 8);
\node[above] at (11.5, 8) {1. Coded Step};

\draw[black, fill = white!20!black] (16, 11) rectangle (26, 13);
\draw[black, fill = white!40!black] (18, 11) rectangle (19, 13);
\node[right] at (26, 12) {${\cal L}_1$};

\node at (21, 10.25) {$\vdots$};

\draw[black, fill = white!20!black] (16, 7) rectangle (26, 9);
\draw[black, fill = white!80!black] (18, 7) rectangle (19, 9);
\node[right] at (26, 8) {${\cal L}_i$};

\node at (21, 6.25) {$\vdots$};

\draw[black, fill = white!20!black] (16, 3) rectangle (26, 5);
\draw[black, fill = white!60!black] (18, 3) rectangle (19, 5);
\node[right] at (26, 4) {${\cal L}_K$};

\draw[black, fill = white!20!black] (-4.5, -3) rectangle (5.5, -1);
\draw[black, fill = white!80!black] (-2.5, -3) rectangle (-1.5, -1);
\node[right] at (5.5, -2) {${\cal L}_j$};
\draw [->] (8, -2) -- (15, -2);
\node[above] at (11.5, -2) {2. Sieving Step};

\draw[black, fill = white!20!black] (16, -3) rectangle (26, -1);
\draw[black, fill = white!80!black] (16, -3) rectangle (19, -1);
\node[right] at (26, -2) {${\cal S}_j$};

\node at (-4, 8.5) {1. $\norm{(\mathbf{a_1} - \mathbf{a_2})_{[N_{i-1}+1:N_i]}} < B\sqrt{n_i}$};

\node at (-4, 4.5) {2. $\norm{\mathbf{a}_{[1:N_i]}} < B\sqrt{N_i}$};

\end{tikzpicture}
\end{center}
\caption{A micro picture of how one step of coded-BKW with sieving works. Slightly changed version of Figure 1 from \cite{FullCodedBKWWithSieving}. Each sample gets mapped to one list ${\cal L}_j$ out of $K$ lists. Sieving is then applied to each list to form new lists ${\cal S}_j$.}
\label{fig:micro}
\end{figure*}

}

\subsubsection{Using Different Reduction Factors}
In \cite{FullCodedBKWWithSieving} the idea of finding an optimal reduction factor was introduced. Instead of making sure the $N_i$ positions currently considered are as small as the $N_{i - 1}$ positions in the previous step, they are made to be $\gamma$ times as large. Depending on the parameter setting the optimal strategy is either to use $\gamma < 1$ or $\gamma > 1$, or in other words to gradually decrease or increase the values in the $\mathbf{a}$ vector. The final average magnitude is still less than the same constant $B$.

The original coded-BKW with sieving algorithm from \cite{AC:GJMS17} is the special case where $\gamma = 1$ and coded-BKW is the special case where $\gamma = \sqrt{2}$.

For an illustration of how the different BKW algorithms reduce the $\mathbf{a}$ vector, see Figure \ref{fig:highlevel}.

\iftoggle{color}{%
	\definecolor{myred}{rgb}{1, 0.25, 0.21}
\definecolor{myblue}{rgb}{0, 0.45, 0.85}

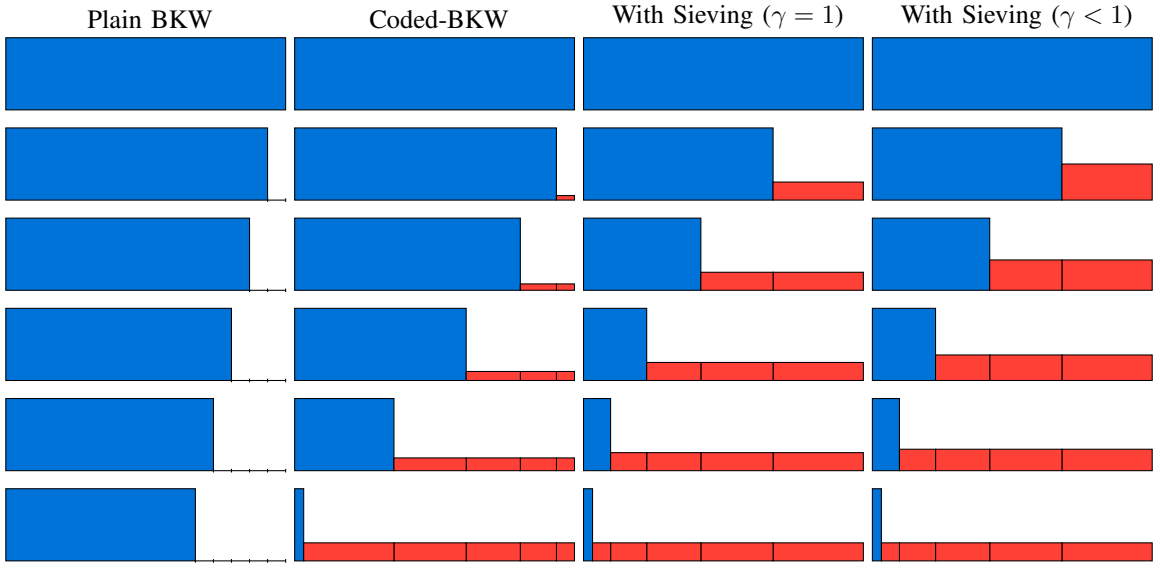
\begin{figure*}[t]
\begin{center}
\begin{tikzpicture}[scale = 0.24]


\draw[black, fill = myblue] (-0.5, 0) rectangle (15, 4);
\node [above] at (7.5, 4) {Plain BKW};

\draw[black, fill = myblue] (15.5, 0) rectangle (31, 4);
\node [above] at (23.5, 4) {Coded-BKW};

\draw[black, fill = myblue] (31.5, 0) rectangle (47, 4);
\node [above] at (39.5, 4) {With Sieving ($\gamma = 1$)};

\draw[black, fill = myblue] (47.5, 0) rectangle (63, 4);
\node [above] at (55.5, 4) {With Sieving ($\gamma < 1$)};


\draw[black, fill = myblue] (-0.5, -5) rectangle (14, -1);
\draw[black, fill = myblue] (-0.5, -10) rectangle (13, -6);
\draw[black, fill = myblue] (-0.5, -15) rectangle (12, -11);
\draw[black, fill = myblue] (-0.5, -20) rectangle (11, -16);
\draw[black, fill = myblue] (-0.5, -25) rectangle (10, -21);

\draw[] (14, -5) -- (15, -5);
\draw[] (14, -5.1) -- (14, -4.9);
\draw[] (15, -5.1) -- (15, -4.9);

\draw[] (13, -10) -- (15, -10);
\draw[] (13, -10.1) -- (13, -9.9);
\draw[] (14, -10.1) -- (14, -9.9);
\draw[] (15, -10.1) -- (15, -9.9);

\draw[] (12, -15) -- (15, -15);
\draw[] (12, -15.1) -- (12, -14.9);
\draw[] (13, -15.1) -- (13, -14.9);
\draw[] (14, -15.1) -- (14, -14.9);
\draw[] (15, -15.1) -- (15, -14.9);

\draw[] (11, -20) -- (15, -20);
\draw[] (11, -20.1) -- (11, -19.9);
\draw[] (12, -20.1) -- (12, -19.9);
\draw[] (13, -20.1) -- (13, -19.9);
\draw[] (14, -20.1) -- (14, -19.9);
\draw[] (15, -20.1) -- (15, -19.9);

\draw[] (10, -25) -- (15, -25);
\draw[] (10, -25.1) -- (10, -24.9);
\draw[] (11, -25.1) -- (11, -24.9);
\draw[] (12, -25.1) -- (12, -24.9);
\draw[] (13, -25.1) -- (13, -24.9);
\draw[] (14, -25.1) -- (14, -24.9);
\draw[] (15, -25.1) -- (15, -24.9);


\draw[black, fill = myblue] (15.5, -5) rectangle (30, -1);
\draw[black, fill = myblue] (15.5, -10) rectangle (28, -6);
\draw[black, fill = myblue] (15.5, -15) rectangle (25, -11);
\draw[black, fill = myblue] (15.5, -20) rectangle (21, -16);
\draw[black, fill = myblue] (15.5, -25) rectangle (16, -21);

\draw[black, fill = myred] (30, -5) rectangle (31, -5 + .25);

\draw[black, fill = myred] (30, -10) rectangle (31, -10 + 0.3536);
\draw[black, fill = myred] (28, -10) rectangle (30, -10 + 0.3536);

\draw[black, fill = myred] (30, -15) rectangle (31, -15 + 0.5);
\draw[black, fill = myred] (28, -15) rectangle (30, -15 + 0.5);
\draw[black, fill = myred] (25, -15) rectangle (28, -15 + 0.5);

\draw[black, fill = myred] (30, -20) rectangle (31, -20 + 0.7071);
\draw[black, fill = myred] (28, -20) rectangle (30, -20 + 0.7071);
\draw[black, fill = myred] (25, -20) rectangle (28, -20 + 0.7071);
\draw[black, fill = myred] (21, -20) rectangle (25, -20 + 0.7071);

\draw[black, fill = myred] (30, -25) rectangle (31, -25 + 1);
\draw[black, fill = myred] (28, -25) rectangle (30, -25 + 1);
\draw[black, fill = myred] (25, -25) rectangle (28, -25 + 1);
\draw[black, fill = myred] (21, -25) rectangle (25, -25 + 1);
\draw[black, fill = myred] (16, -25) rectangle (21, -25 + 1);


\draw[black, fill = myblue] (31.5, -5) rectangle (42, -1);
\draw[black, fill = myblue] (31.5, -10) rectangle (38, -6);
\draw[black, fill = myblue] (31.5, -15) rectangle (35, -11);
\draw[black, fill = myblue] (31.5, -20) rectangle (33, -16);
\draw[black, fill = myblue] (31.5, -25) rectangle (32, -21);

\draw[black, fill = myred] (42, -5) rectangle (47, -5 + 1);

\draw[black, fill = myred] (42, -10) rectangle (47, -10 + 1);
\draw[black, fill = myred] (38, -10) rectangle (42, -10 + 1);

\draw[black, fill = myred] (42, -15) rectangle (47, -15 + 1);
\draw[black, fill = myred] (38, -15) rectangle (42, -15 + 1);
\draw[black, fill = myred] (35, -15) rectangle (38, -15 + 1);

\draw[black, fill = myred] (42, -20) rectangle (47, -20 + 1);
\draw[black, fill = myred] (38, -20) rectangle (42, -20 + 1);
\draw[black, fill = myred] (35, -20) rectangle (38, -20 + 1);
\draw[black, fill = myred] (33, -20) rectangle (35, -20 + 1);

\draw[black, fill = myred] (42, -25) rectangle (47, -25 + 1);
\draw[black, fill = myred] (38, -25) rectangle (42, -25 + 1);
\draw[black, fill = myred] (35, -25) rectangle (38, -25 + 1);
\draw[black, fill = myred] (33, -25) rectangle (35, -25 + 1);
\draw[black, fill = myred] (32, -25) rectangle (33, -25 + 1);


\draw[black, fill = myblue] (47.5, -5) rectangle (58, -1);
\draw[black, fill = myblue] (47.5, -10) rectangle (54, -6);
\draw[black, fill = myblue] (47.5, -15) rectangle (51, -11);
\draw[black, fill = myblue] (47.5, -20) rectangle (49, -16);
\draw[black, fill = myblue] (47.5, -25) rectangle (48, -21);

\draw[black, fill = myred] (58, -5) rectangle (63, -5 + 2);
\draw[black, fill = myred] (58, -10) rectangle (63, -10 + 2 * 0.840896);
\draw[black, fill = myred] (54, -10) rectangle (58, -10 + 2 * 0.840896);
\draw[black, fill = myred] (58, -15) rectangle (63, -15 + 2 * 0.840896^2);
\draw[black, fill = myred] (54, -15) rectangle (58, -15 + 2 * 0.840896^2);
\draw[black, fill = myred] (51, -15) rectangle (54, -15 + 2 * 0.840896^2);
\draw[black, fill = myred] (58, -20) rectangle (63, -20 + 2 * 0.840896^3);
\draw[black, fill = myred] (54, -20) rectangle (58, -20 + 2 * 0.840896^3);
\draw[black, fill = myred] (51, -20) rectangle (54, -20 + 2 * 0.840896^3);
\draw[black, fill = myred] (49, -20) rectangle (51, -20 + 2 * 0.840896^3);
\draw[black, fill = myred] (58, -25) rectangle (63, -25 + 1);
\draw[black, fill = myred] (54, -25) rectangle (58, -25 + 1);
\draw[black, fill = myred] (51, -25) rectangle (54, -25 + 1);
\draw[black, fill = myred] (49, -25) rectangle (51, -25 + 1);
\draw[black, fill = myred] (48, -25) rectangle (49, -25 + 1);

\end{tikzpicture}
\end{center}
\caption{A high-level illustration of how the different versions of the BKW algorithm work. The $x$-axis represents positions in the $\mathbf{a}$ vector, and the $y$-axis depicts the average absolute value of the corresponding position. The blue color corresponds to positions that have not been reduced yet and the red color corresponds to reduced positions. The last few positions are used for guessing. The figure is a modified version of Figures 3 and 8 from \cite{FullCodedBKWWithSieving}.}
\label{fig:highlevel}
\end{figure*}
}{
\begin{figure*}[t]
\begin{center}
\begin{tikzpicture}[scale = 0.24]


\draw[fill = white!66!black] (-0.5, 0) rectangle (15, 4);
\node [above] at (7.5, 4) {Plain BKW};

\draw[fill = white!66!black] (15.5, 0) rectangle (31, 4);
\node [above] at (23.5, 4) {Coded-BKW};

\draw[black, fill = white!66!black] (31.5, 0) rectangle (47, 4);
\node [above] at (39.5, 4) {With Sieving ($\gamma = 1$)};

\draw[black, fill = white!66!black] (47.5, 0) rectangle (63, 4);
\node [above] at (55.5, 4) {With Sieving ($\gamma < 1$)};


\draw[fill = white!66!black] (-0.5, -5) rectangle (14, -1);
\draw[fill = white!66!black] (-0.5, -10) rectangle (13, -6);
\draw[fill = white!66!black] (-0.5, -15) rectangle (12, -11);
\draw[fill = white!66!black] (-0.5, -20) rectangle (11, -16);
\draw[fill = white!66!black] (-0.5, -25) rectangle (10, -21);

\draw[] (14, -5) -- (15, -5);
\draw[] (14, -5.1) -- (14, -4.9);
\draw[] (15, -5.1) -- (15, -4.9);

\draw[] (13, -10) -- (15, -10);
\draw[] (13, -10.1) -- (13, -9.9);
\draw[] (14, -10.1) -- (14, -9.9);
\draw[] (15, -10.1) -- (15, -9.9);

\draw[] (12, -15) -- (15, -15);
\draw[] (12, -15.1) -- (12, -14.9);
\draw[] (13, -15.1) -- (13, -14.9);
\draw[] (14, -15.1) -- (14, -14.9);
\draw[] (15, -15.1) -- (15, -14.9);

\draw[] (11, -20) -- (15, -20);
\draw[] (11, -20.1) -- (11, -19.9);
\draw[] (12, -20.1) -- (12, -19.9);
\draw[] (13, -20.1) -- (13, -19.9);
\draw[] (14, -20.1) -- (14, -19.9);
\draw[] (15, -20.1) -- (15, -19.9);

\draw[] (10, -25) -- (15, -25);
\draw[] (10, -25.1) -- (10, -24.9);
\draw[] (11, -25.1) -- (11, -24.9);
\draw[] (12, -25.1) -- (12, -24.9);
\draw[] (13, -25.1) -- (13, -24.9);
\draw[] (14, -25.1) -- (14, -24.9);
\draw[] (15, -25.1) -- (15, -24.9);


\draw[fill = white!66!black] (15.5, -5) rectangle (30, -1);
\draw[fill = white!66!black] (15.5, -10) rectangle (28, -6);
\draw[fill = white!66!black] (15.5, -15) rectangle (25, -11);
\draw[fill = white!66!black] (15.5, -20) rectangle (21, -16);
\draw[fill = white!66!black] (15.5, -25) rectangle (16, -21);

\draw[fill = white!45!black] (30, -5) rectangle (31, -5 + .25);

\draw[fill = white!45!black] (30, -10) rectangle (31, -10 + 0.3536);
\draw[fill = white!45!black] (28, -10) rectangle (30, -10 + 0.3536);

\draw[fill = white!45!black] (30, -15) rectangle (31, -15 + 0.5);
\draw[fill = white!45!black] (28, -15) rectangle (30, -15 + 0.5);
\draw[fill = white!45!black] (25, -15) rectangle (28, -15 + 0.5);

\draw[fill = white!45!black] (30, -20) rectangle (31, -20 + 0.7071);
\draw[fill = white!45!black] (28, -20) rectangle (30, -20 + 0.7071);
\draw[fill = white!45!black] (25, -20) rectangle (28, -20 + 0.7071);
\draw[fill = white!45!black] (21, -20) rectangle (25, -20 + 0.7071);

\draw[fill = white!45!black] (30, -25) rectangle (31, -25 + 1);
\draw[fill = white!45!black] (28, -25) rectangle (30, -25 + 1);
\draw[fill = white!45!black] (25, -25) rectangle (28, -25 + 1);
\draw[fill = white!45!black] (21, -25) rectangle (25, -25 + 1);
\draw[fill = white!45!black] (16, -25) rectangle (21, -25 + 1);


\draw[fill = white!66!black] (31.5, -5) rectangle (42, -1);
\draw[fill = white!66!black] (31.5, -10) rectangle (38, -6);
\draw[fill = white!66!black] (31.5, -15) rectangle (35, -11);
\draw[fill = white!66!black] (31.5, -20) rectangle (33, -16);
\draw[fill = white!66!black] (31.5, -25) rectangle (32, -21);

\draw[fill = white!45!black] (42, -5) rectangle (47, -5 + 1);

\draw[fill = white!45!black] (42, -10) rectangle (47, -10 + 1);
\draw[fill = white!45!black] (38, -10) rectangle (42, -10 + 1);

\draw[fill = white!45!black] (42, -15) rectangle (47, -15 + 1);
\draw[fill = white!45!black] (38, -15) rectangle (42, -15 + 1);
\draw[fill = white!45!black] (35, -15) rectangle (38, -15 + 1);

\draw[fill = white!45!black] (42, -20) rectangle (47, -20 + 1);
\draw[fill = white!45!black] (38, -20) rectangle (42, -20 + 1);
\draw[fill = white!45!black] (35, -20) rectangle (38, -20 + 1);
\draw[fill = white!45!black] (33, -20) rectangle (35, -20 + 1);

\draw[fill = white!45!black] (42, -25) rectangle (47, -25 + 1);
\draw[fill = white!45!black] (38, -25) rectangle (42, -25 + 1);
\draw[fill = white!45!black] (35, -25) rectangle (38, -25 + 1);
\draw[fill = white!45!black] (33, -25) rectangle (35, -25 + 1);
\draw[fill = white!45!black] (32, -25) rectangle (33, -25 + 1);


\draw[black, fill = white!66!black] (47.5, -5) rectangle (58, -1);
\draw[black, fill = white!66!black] (47.5, -10) rectangle (54, -6);
\draw[black, fill = white!66!black] (47.5, -15) rectangle (51, -11);
\draw[black, fill = white!66!black] (47.5, -20) rectangle (49, -16);
\draw[black, fill = white!66!black] (47.5, -25) rectangle (48, -21);

\draw[black, fill = white!45!black] (58, -5) rectangle (63, -5 + 2);
\draw[black, fill = white!45!black] (58, -10) rectangle (63, -10 + 2 * 0.840896);
\draw[black, fill = white!45!black] (54, -10) rectangle (58, -10 + 2 * 0.840896);
\draw[black, fill = white!45!black] (58, -15) rectangle (63, -15 + 2 * 0.840896^2);
\draw[black, fill = white!45!black] (54, -15) rectangle (58, -15 + 2 * 0.840896^2);
\draw[black, fill = white!45!black] (51, -15) rectangle (54, -15 + 2 * 0.840896^2);
\draw[black, fill = white!45!black] (58, -20) rectangle (63, -20 + 2 * 0.840896^3);
\draw[black, fill = white!45!black] (54, -20) rectangle (58, -20 + 2 * 0.840896^3);
\draw[black, fill = white!45!black] (51, -20) rectangle (54, -20 + 2 * 0.840896^3);
\draw[black, fill = white!45!black] (49, -20) rectangle (51, -20 + 2 * 0.840896^3);
\draw[black, fill = white!45!black] (58, -25) rectangle (63, -25 + 1);
\draw[black, fill = white!45!black] (54, -25) rectangle (58, -25 + 1);
\draw[black, fill = white!45!black] (51, -25) rectangle (54, -25 + 1);
\draw[black, fill = white!45!black] (49, -25) rectangle (51, -25 + 1);
\draw[black, fill = white!45!black] (48, -25) rectangle (49, -25 + 1);

\end{tikzpicture}
\end{center}
\caption{A high-level illustration of how the different versions of the BKW algorithm work. The $x$-axis represents positions in the $\mathbf{a}$ vector, and the $y$-axis depicts the average absolute value of the corresponding position. The lighter color corresponds to positions that have not been reduced yet and the darker color corresponds to reduced positions. The last few positions are used for guessing. The figure is a modified version of Figures 3 and 8 from \cite{FullCodedBKWWithSieving}.}
\label{fig:highlevel}
\end{figure*}
}

\section{Coded-BKW with Sieving with Increasing Reduction Factors}
\label{sec:NewAlgorithm}

The new idea in this paper is to use different reduction factors $\gamma_i$ in different steps $i$. The idea is that in the earlier steps the sieving is cheap, and we can therefore use small values of $\gamma_i$. Gradually the sieving procedure gets more and more expensive, forcing us to increase the value of $\gamma_i$.

Assume that we take $t_2$ steps of coded-BKW with sieving in total and let $\gamma_1 = \gamma_s$ and$\gamma_{t_2} = \gamma_f$. We ended up choosing an arithmetic progression, that is we let

\begin{equation*}
	\gamma_i = \gamma_s + \frac{\gamma_f - \gamma_s}{t_2 - 1} (i - 1).
\end{equation*}

We also tried a geometric and logarithmic progression of the $\gamma_i$ values, both leading to a slightly worse complexity. A power progression lead to an expression that had to be estimated numerically and resulted in almost exactly the same results as the arithmetic progression.

\section{Asymptotic Complexity}
\label{sec:Asymptotics}

Asymptotically we let $q = n^{c_q}$ and $\sigma = n^{c_s}$, where $c_q$ and $c_s$ are constants. For most algorithms and settings the asymptotic complexity of solving LWE is $2^{cn + o(n)}$, where the exponent $c$ depends on $c_q$ and $c_s$. We leave settings such as a binary secret or a superpolynomial $q$ for future research.

We will now quickly cover the asymptotic complexities of the Arora-Ge algorithm, lattice-based algorithms and all the previous versions of BKW, as a function of $c_q$ and $c_s$. Initially, the assumed setting is one with a classical computer and an exponential amount of samples. Other settings will be discussed later.

\subsubsection{Complexity Exponent for Lattice Sieving}
The value $\lambda(\gamma)$ for lattice sieving using a reduction factor $\gamma$ is the best available complexity exponent for doing lattice sieving. It is (currently) calculated by doing the optimization from the section about the total cost of sieving in \cite{SODA:BDGL16}, replacing the angle $\pi/3$ by $\theta = 2\arcsin(\gamma/2)$ and replacing $N = (4/3)^{n/2}$ by $(1/\sin(\theta))^n$. For $\gamma = 1$ we get $\lambda \approx 0.292$.

\subsubsection{Quantum Setting}
If having access to a quantum computer, Grover's algorithm can be used to speed up the lattice sieving, see \cite{GroverSieving}, resulting in slightly improved complexity exponents. For $\gamma = 1$ we get $\lambda \approx 0.265$.

\subsection{Arora-Ge and Lattice-based Methods}
The Arora-Ge algorithm is polynomial when $c_s < 0.5$ and superexponential when $c_s > 0.5$, making it viable if and only if $c_s$ is too small for Regev's reduction proof to apply \cite{STOC:Regev05}. Lattice-based algorithms can solve LWE with a time complexity exponent of $2 \lambda c_q/(c_q - c_s + 1/2)^2$, using an exponential amount of memory \cite{AsymptoticLWE}. 

\subsection{Plain and Coded BKW}
The time and space complexity for solving LWE using plain BKW is $c_q/(2(c_q - c_s) + 1)$ \cite{Albrecht2015}, and using Coded-BKW is $(1/c_q + 2\ln(c_q/c_s))^{-1}$ \cite{C:KirFou15}.

\subsection{Coded-BKW with sieving}
The time (and space) complexity of solving the LWE problem using coded-BKW with sieving gets calculated by solving increasingly difficult optimization problems. In both Theorem \ref{thm:ConstantGamma} and \ref{thm:VaryingGamma}, the parameter $\alpha$ decides how large part of the samples should be pre-processed with plain BKW steps.

\begin{thm}
\label{thm:ConstantGamma}
The time and space complexity of coded-BKW with sieving and a constant value of the reduction factor $\gamma$, is $2^{cn + o(n)}$, where $c$ is the solution to the following optimization problem.

  \begin{equation*}
  \begin{aligned}
  & \underset{\alpha,\gamma}{\textbf{minimize}}
  & & c(\alpha,\gamma) =
  \Bigg( \frac{2(c_q - c_s) + 1 - \alpha}{c_q} + \\
  &&& \frac{1}{\lambda(\gamma)} \left(  1 - \frac{c_{s}}{\alpha \log_2 \gamma + c_{s}} \cdot \exp(I(\alpha, \gamma)) \right)     \Bigg)^{-1} \\
  & \textbf{subject to}
  & & 0 \leq \alpha \leq 2(c_{q} - c_{s}) + 1, \\
  &&& 0 < \gamma \leq \sqrt{2}.
  \end{aligned}
  \end{equation*}

Here, we have

\begin{equation*}
I(\alpha, \gamma) = \int_{0}^{\alpha} \frac{\log_2 \gamma - \lambda(\gamma)}{t\log_2 \gamma + c_s} dt.
\end{equation*}

By setting $\gamma = 1$ we get (a restatement of) the complexity of the original coded-BKW with sieving algorithm \cite{AC:GJMS17}.

\end{thm}
  
\begin{IEEEproof}
The theorem is a slight restatement of Theorem 7 from \cite{FullCodedBKWWithSieving}, to make it more similar to Theorem \ref{thm:VaryingGamma} of this paper. Theorem 7 from \cite{FullCodedBKWWithSieving} includes both the proof and the underlying heuristic assumptions the proof is based on.
\end{IEEEproof}

\subsection{Coded-BKW with Sieving with Increasing Reduction Factors}
The complexity of the new algorithm is covered in the following theorem.

\begin{thm}
\label{thm:VaryingGamma}
The time and space complexity of coded-BKW with sieving and an arithmetic progression of the $\gamma_i$ values, is $2^{cn + o(n)}$, where $c$ is the solution to the following optimization problem.

\begin{equation*}
\begin{aligned}
\footnotesize
& \underset{\alpha,\gamma_s, \gamma_f}{\textbf{minimize}}
& & c(\alpha,\gamma_s, \gamma_f) =
\bigg( \frac{2(c_q - c_s) + 1 - \alpha}{c_q} + \\
&&& \int_{0}^{\alpha} \frac{c_s}{\left(t \cdot \ell (t) + c_s\right)^2} \cdot \exp(I(t; \alpha, \gamma_s, \gamma_f)) dt     \bigg)^{-1} \\
& \textbf{subject to}
& & 0 \leq \alpha \leq 2(c_{q} - c_{s}) + 1, \\
&&& 0 < \gamma_s < \gamma_f \leq \sqrt{2}. \\
\end{aligned}
\end{equation*}

Here, we have

\begin{equation*}
I(t; \alpha, \gamma_s, \gamma_f) = \int_{0}^{t} \frac{\log_2 \gamma(s) - \lambda(\gamma(s))}{s \cdot \ell(s) + c_s} ds,	
\end{equation*}

and

\begin{align*}
	\gamma(s) 	&= \gamma_s + \frac{\alpha - s}{\alpha} (\gamma_f - \gamma_s), \\
	\ell(s)		&= \left( \frac{\gamma_f \ln(\gamma_f) - \gamma(t) \ln(\gamma(t))}{\gamma_f - \gamma_s} \frac{\alpha}{s} - 1 \right)/\ln(2).
\end{align*}
	
\end{thm}

It should be mentioned that the objetive function of the optimization problem here would change slightly if another method for the progression of the $\gamma_i$ values was chosen.

\begin{IEEEproof}
\iftoggle{full}{%
	A proof of Theorem \ref{thm:VaryingGamma} can be found in the appendix.
}{
	The proof of Theorem \ref{thm:VaryingGamma} is omitted here due to the page restriction. A proof can be found in the full version.
}
\end{IEEEproof}


\subsection{Polynomial Number of Samples}
With access to only a polynomial number of samples the complexity exponent of lattice-based algorithms changes to $2 \lambda c_q/(c_q - c_s)^2$ \cite{AsymptoticLWE}.

When using BKW with access to only a polynomial number of samples, amplification is used to increase the number of samples, at the cost of an increased noise. For plain and coded BKW the complexity exponents change to $c_q/(2(c_q - c_s))$ and $(1/c_q + 2\ln(c_q/c_s))^{-1}$ \cite{AsymptoticLWE}.

In the optimization problems in Theorem \ref{thm:ConstantGamma} and \ref{thm:VaryingGamma} the upper limit of $\alpha$ changes to $2(c_q - c_s)$. For each theorem the numerator in the first term of the objective function changes to $2(c_q - c_s) - \alpha$.

\section{Results}
\label{sec:Results}

Let us use the Regev instances as a case study, in other words, let $c_q = 2$ and $c_s = 1.5$ \cite{STOC:Regev05}. Table \ref{tab:Regev} shows the complexity exponent for coded-BKW with sieving, with $\gamma = 1$, an optimized constant $\gamma$ and an arithmetic progression of the $\gamma$ values, for four different scenarios. Either we use classical or quantum computers and either we have access to a polynomial or an exponential number of samples. Notice how using increasing reduction factors improves the complexity exponent in all the scenarios.

\begin{figure}[!h]
	\centering
	\iftoggle{color}{%
		\includegraphics[width = 0.49 \textwidth]{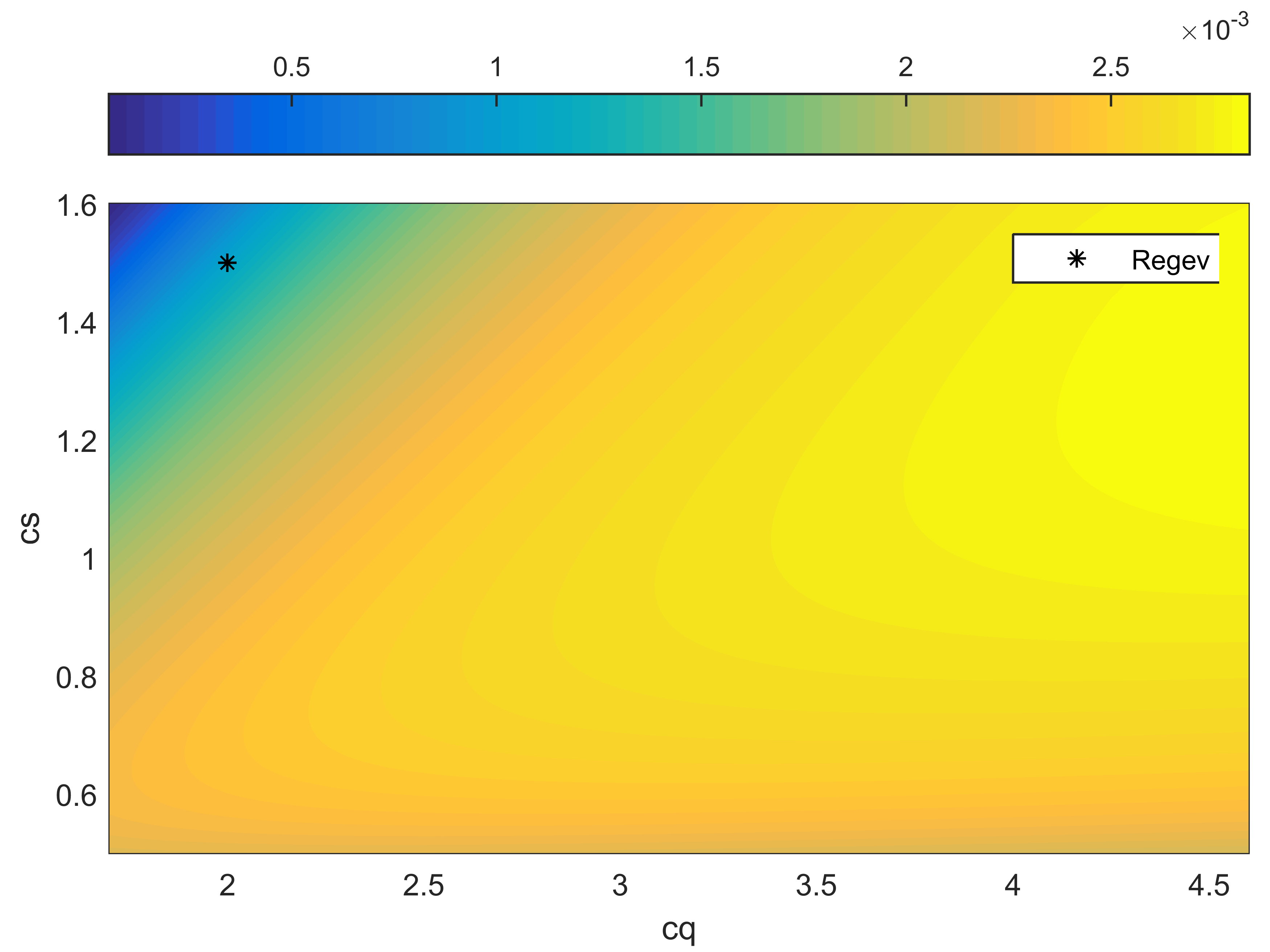}
	}{
		\includegraphics[width = 0.49 \textwidth]{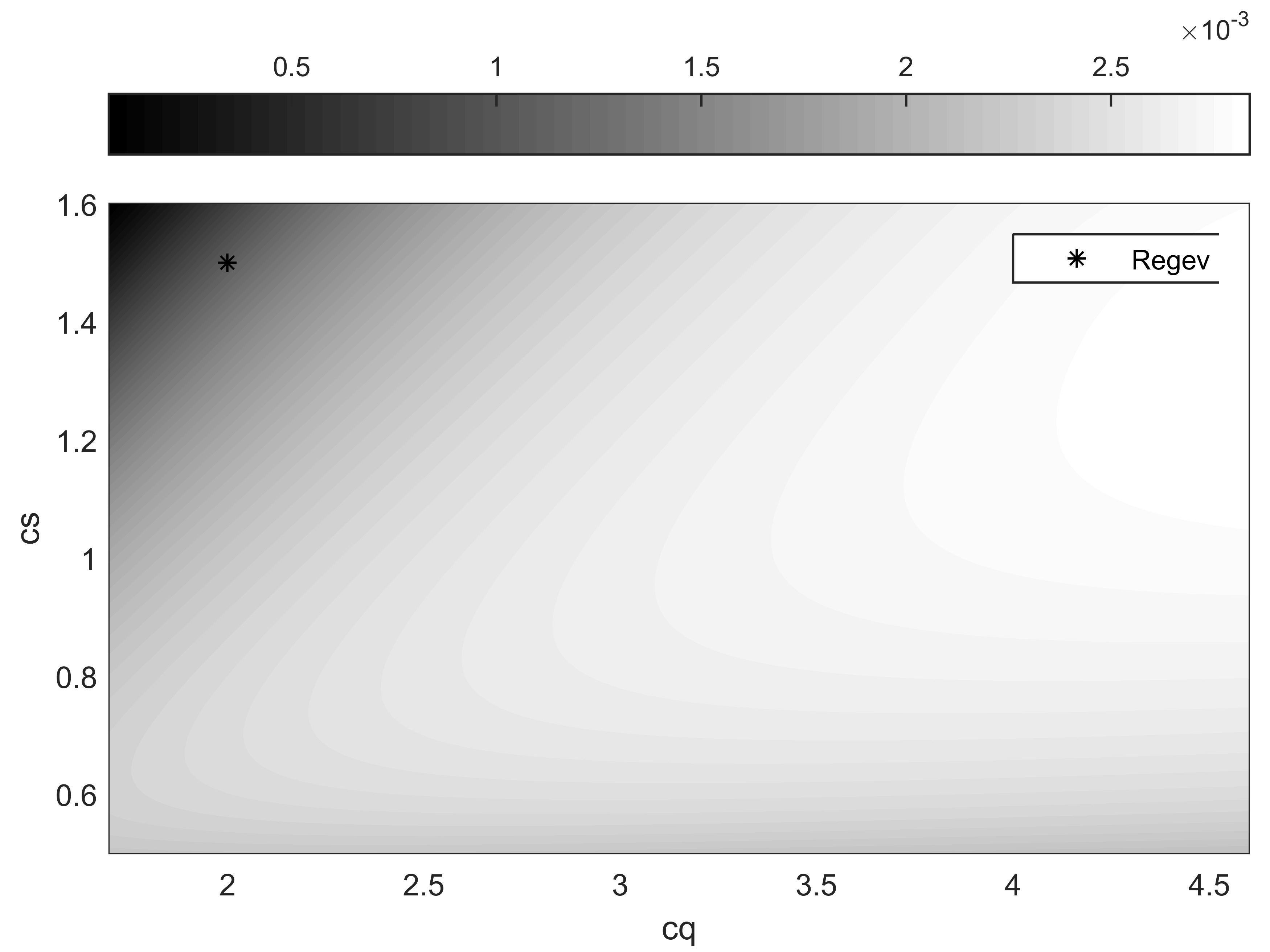}
	}
	
	\caption{The improvement in the complexity exponent when going from a constant $\gamma$ to an arithmetic progression of the $\gamma$ values}
	\label{fig:ConstantVSArithmetic}
\end{figure}

In all the scenarios in the Regev setting BKW algorithms beat lattice-based algorithms. For a picture comparing the asymptotic complexity of the different BKW versions with lattice-based algorithms in other settings, see Figure 5 and 7 in \cite{FullCodedBKWWithSieving}. The new version constitutes an improvement compared to the constant $\gamma$ algorithm for all parameter pairs $(c_q, c_s)$, as can be seen in Figure \ref{fig:ConstantVSArithmetic}.

The source code used for calculating all the complexity exponents in the different scenarios can be found on GitHub\footnote{\url{https://github.com/ErikMaartensson/BKWIncreasingReductionFactors}}.


\begin{table}[h]
	\caption{\label{tab:Regev} The asymptotic complexity exponent for the different versions of BKW in the Regev setting.}
	\centering
	\sffamily \begin{tabularx}{1\linewidth}{>{\centering\arraybackslash}m{1cm}|>{\centering\arraybackslash}m{3.1cm}|>{\centering\arraybackslash}m{3.1cm}}
		
		& Classical & Quantum \\
		& Setting & Setting \\
		\hline
		\begin{sideways}\parbox{2cm}{\centering Exponential samples} \end{sideways} &
		$
			\begin{aligned}
				&0.8951\text{ } (\gamma = 1) \\
				&0.8927\text{ } (\gamma \text{ constant}) \\
				&\mathbf{0.8917}\text{ } (\gamma \text{ arithmetic})
			\end{aligned}
		$ &
		$
		\begin{aligned}
			&0.8856\text{ } (\gamma=1) \\
			&0.8795\text{ } (\gamma \text{ constant}) \\
			&\mathbf{0.8782}\text{ } (\gamma \text{ arithmetic})
		\end{aligned}
		$
		\\
		\hline
		
		\begin{sideways}\parbox{2cm}{\centering Polynomial samples} \end{sideways} &
		$
		\begin{aligned}
			&1.6507\text{ } (\gamma=1) \\
			&1.6417\text{ } (\gamma \text{ constant}) \\
			&\mathbf{1.6399}\text{ } (\gamma \text{ arithmetic})
		\end{aligned}
		$
		&
		$
		\begin{aligned}
			&1.6364\text{ } (\gamma=1) \\
			&1.6211\text{ } (\gamma \text{ constant}) \\
			&\mathbf{1.6168}\text{ } (\gamma \text{ arithmetic})
		\end{aligned}
		$
		\\
		\hline
	\end{tabularx} \normalfont
	
\end{table}

\section{Conclusions}
\label{sec:Conclusions}

We have developed a new version of coded-BKW with sieving, achieving a further improved asymptotic complexity. We have also improved the complexity when having access to a quantum computer or only having access to a polynomial number of samples. All BKW algorithms solve a modified version of the 2-list problem, where we also have a modular reduction and have a limited number of total steps. Generalizing the optimization of the BKW algorithm from this perspective is an interesting new research idea. Another possible further research direction is looking at the complexity of different versions of BKW when only having access to a limited amount of memory, like was briefly started in \cite{C:EHKMS18}. Finally, it is interesting to investigate the concrete complexity of coded-BKW with sieving and implement it to see when BKW starts to beat lattice-type algorithms in practise.

\section*{Acknowledgment}
The author would like to acknowledge Qian Guo, whose idea of using a reduction factor $\gamma \neq 1$ this paper generalizes. The author would also like to thank Thomas Johansson and Paul Stankovski Wagner for fruitful discussions during the writing of this paper.


\bibliography{abbrev0,crypto,other}
\bibliographystyle{IEEEtran}


\iftoggle{full}{%
	\newpage
\appendix

Let us prove Theorem \ref{thm:VaryingGamma}.
\begin{IEEEproof}
This proof is generalization of the proof of Theorem \ref{thm:ConstantGamma}, as proven in \cite{FullCodedBKWWithSieving}. The proof structure is similar, but the proof is also much longer. To avoid making it longer than necessary, some notation is borrowed from the proof in \cite{FullCodedBKWWithSieving}.

Instead of using a constant reduction factor $\gamma$, we use a low reduction factor in the first steps when the sieving is cheap, and then gradually increase the reduction factor. Let $\gamma_i$ denote the reduction factor in step $i$. We have the constraints

\begin{equation*}
0 < \gamma_1 < \gamma_2 < \ldots < \gamma_{t_2} \leq \sqrt{2}.
\end{equation*}

Let $\lambda_i = \lambda(\gamma_i)$ denote the corresponding complexity exponent for nearest neighbor searching using the LSF algorithm. By $\log$ we denote $\log_2$.

Later we will let $\gamma_i$ increase arithmetically. Other progressions also work, and as long as possible we will use a general progression of the $\gamma_i$ values.

We start by performing $t_{1}$ plain BKW steps and then $t_{2}$ coded-BKW with sieving steps with parameters $(\lambda_i, \gamma_i)$ in the $i$\textsuperscript{th} of the latter steps. Let $t_{2 } =  \alpha \log n + \mathcal{O}(1)$ and $t_{1} = \beta \log n = (2(c_{q} - c_{s}) + 1 - \alpha) \log n + \mathcal{O}(1)$. We have the constraint that $0 \leq \alpha \leq 2(c_q - c_s) + 1$. Like previously the step size of the plain BKW steps is

\begin{equation*}
b = \frac{cn}{c_{q}\log n}.
\end{equation*} 

Let $n_1$ be the length of the first coded-BKW with sieving step. Let $B_i$ denote the magnitude of the values at step $i$ and $B$ denote the final magnitude of the values. Then we get

\begin{equation*}
B_1 = \frac{B}{\Pi_{i = 1}^{t_2} \gamma_i}.
\end{equation*}

Therefore,

\begin{equation*}
\left(  \log(\Pi_{i = 1}^{t_2} \gamma_i) + c_s\log(n)  \right) \cdot n_1 = cn - \lambda_1\ \cdot n_1.
\end{equation*}

Thus,

\begin{equation}
\begin{split}
n_{1} &= \frac{cn}{c_{s}\log n + \log(\Pi_{i = 1}^{t_2} \gamma_i) + \lambda_1}  \\
\label{eq:n1}
&= \frac{cn}{c_{s}\log n + \log(\Pi_{i = 1}^{t_2} \gamma_i)} \cdot (1 + \Theta(\log^{-1} n)).
\end{split}
\end{equation}

Analogously with step 1, for step $i$ we get

\begin{equation}
\label{eq:stepJ}
\left(  \log(\Pi_{j = i}^{t_2} \gamma_j) + c_s\log(n)  \right) \cdot n_i = cn - \lambda_i\ \sum_{j=1}^i n_j.
\end{equation}

Use $\star_i$ to denote the expression within the outer parantheses in the left hand-side of \eqref{eq:stepJ}, for step $i$. Multiply \eqref{eq:stepJ} for step $i$ by $\lambda_{i - 1}$, for step $i - 1$ by $\lambda_i$ and subtract the expressions to get

\begin{equation}
\label{eq:Diff}
\lambda_{i - 1} \star_i n_i - \lambda_i \star_{i - 1} n_{i - 1} = cn(\lambda_{i - 1} - \lambda_i) - \lambda_{i - 1} \lambda_i n_i.
\end{equation}

Solving \eqref{eq:Diff} for $n_i$ gives

\begin{equation}
\label{eq:ni}
n_i = \frac{\lambda_i \star_{i - 1} n_{i - 1} + cn(\lambda_{i - 1} - \lambda_i)}{\lambda_{i - 1} \star_i + \lambda_{i - 1} \lambda_i}.
\end{equation}

Calculating $n_{t_2}$ from \eqref{eq:ni} gives

\begin{eqnarray}
\label{eq:smallTerm}
&&n_{t_2} = n_1 \prod_{i=2}^{t_2} \frac{\lambda_i \star_{i - 1}}{\lambda_{i - 1}(\star_i + \lambda_i)} \\
&+& cn \sum_{i=2}^{t_2} \frac{\lambda_{i - 1} - \lambda_i}{\lambda_{i - 1}(\star_i + \lambda_i)} \prod_{j = i + 1}^{t_2}  \left(\frac{\lambda_j \star_{j - 1}}{\lambda_{j - 1}(\star_j + \lambda_j)}   \right).
\label{eq:largeTerm}
\end{eqnarray}

Let us next look at the term \eqref{eq:largeTerm}. First we rewrite the product as

\begin{equation*}
	\begin{split}
	\prod_{j = i + 1}^{t_2} &\left(\frac{\lambda_j \star_{j - 1}}{\lambda_{j - 1}(\star_j + \lambda_j)}   \right) = \prod_{j = i + 1}^{t_2} \frac{\lambda_j}{\lambda_{j - 1}}  \left(\frac{\star_{j - 1}}{\star_j + \lambda_j}   \right) \\
	&= \frac{\lambda_{t_2}}{\lambda_i} \prod_{j = i + 1}^{t_2} \left(\frac{\star_j + \log(\gamma_{j - 1}) + \lambda_j - \lambda_j}{\star_j + \lambda_j}   \right) \\
	&= \frac{\lambda_{t_2}}{\lambda_i} \prod_{j = i + 1}^{t_2} \left(1 + \frac{\log(\gamma_{j - 1}) - \lambda_j}{\star_j + \lambda_j}   \right).
	\end{split}	
\end{equation*}

Use $\Pi_i$ to denote the product and $a_i$ to denote $\star_i + \lambda_i$. We can then write the term \eqref{eq:largeTerm} as

\begin{equation}
\begin{split}
cn&\sum_{i=2}^{t_2} \frac{\lambda_{i - 1} - \lambda_i}{\lambda_{i - 1} a_i} \frac{\lambda_{t_2}}{\lambda_i} \Pi_i \\
&= cn\sum_{i=2}^{t_2} \frac{1}{a_i} \frac{\lambda_{t_2}}{\lambda_i} \Pi_i - cn\sum_{i=2}^{t_2} \frac{1}{a_i} \frac{\lambda_{t_2}}{\lambda_{i - 1}} \Pi_i \\
\label{eq:largeTermModified}
&= cn\left( \frac{1}{a_{t_2}} - \frac{1}{a_2} \frac{\lambda_{t_2}}{\lambda_1} \Pi_2 + \sum_{i=2}^{t_2 - 1} \frac{\lambda_{t_2}}{\lambda_i} \left( \frac{\Pi_i}{a_i} - \frac{\Pi_{i + 1}}{a_{i + 1}} \right) \right).
\end{split}
\end{equation}

Next, we write the expression within paranthesis in the sum in \eqref{eq:largeTermModified} as

\begin{equation*}
\begin{split}
&\frac{\Pi_i a_{i + 1} - \Pi_{i + 1}a_i}{a_i a_{i + 1}} \\
&= \frac{\Pi_{i + 1} \left( \left(   1+  \frac{\log (\gamma_{i}) - \lambda_{i + 1}}{\log(\Pi_{j = i + 1}^{t_2} \gamma_j) + c_s\log(n) + \lambda_{i + 1}}   \right)a_{i + 1}  - a_i \right) }{a_i a_{i + 1}}\\
&= \frac{\Pi_{i + 1} \left(a_{i + 1} +  \log (\gamma_{i}) - \lambda_{i + 1} - a_i \right) }{a_i a_{i + 1}}\\ 
&= \frac{\Pi_{i + 1} \left(\lambda_{i + 1} +  \log(\gamma_{i}) - \lambda_{i + 1} - (\log(\gamma_i) + \lambda_i) \right) }{a_i a_{i + 1}} \\
&= -\frac{\lambda_i}{a_i a_{i + 1}} \Pi_{i + 1}.
\end{split}
\end{equation*}

The product \eqref{eq:smallTerm} can be written as

\begin{equation*}
\frac{cn}{a_1} \frac{\lambda_{t_2}}{\lambda_1} \Pi_1.
\end{equation*}

Thus, adding the two parts \eqref{eq:smallTerm} and \eqref{eq:largeTerm} gives

\begin{equation}
\label{eq:SumProduct}
cn\left( \frac{1}{a_{t_2}} - \lambda_{t_2} \sum_{i=1}^{t_2 - 1} \frac{\Pi_{i + 1}}{a_i a_{i + 1}} \right).
\end{equation}

Next, we want to evaluate $\Pi_{i + 1}$, which can be rewritten as

\begin{equation*}
\begin{split}
&\prod_{j = i + 2}^{t_2}  1+  \frac{\log (\gamma_{j - 1}) - \lambda_j}{\log(\Pi_{k = j}^{t_2} \gamma_k) + c_s\log(n) + \lambda_j} \\
&= \exp \left( \ln \left( \prod_{j = i + 2}^{t_2} 1+  \frac{\log (\gamma_{j - 1}) - \lambda_j}{\log(\Pi_{k = j}^{t_2} \gamma_k) + c_s\log(n) + \lambda_j} \right) \right) \\
&= \resizebox{.93\hsize}{!}{$\exp \left( \sum_{j = i + 2}^{t_2} \frac{\log (\gamma_{j - 1}) - \lambda_j}{\log(\Pi_{k = j}^{t_2} \gamma_k) + c_s\log(n) + \lambda_j} + \Theta \left(\log^{-2} n\right) \right).$}
\end{split}
\end{equation*}

Now the progression of the $\gamma_i$ values needs to be specified. Use an arithmetic progression from $\gamma_1 = \gamma_s$ up to $\gamma_{t_2} = \gamma_f$, where $0 < \gamma_s < \gamma_f \leq \sqrt{2}$. That is, let

\begin{equation*}
\gamma_i = \gamma_s + d(i - 1) = \gamma_s + \frac{\gamma_f - \gamma_s}{t_2 - 1} (i - 1).
\end{equation*}

The idea now is to let $n$ go towards infinity and let the sum in \eqref{eq:SumProduct} approach an integral. If we let $n$ go towards infinity and make a change of variables we get

\begin{equation*}
\left[
\begin{aligned}
t  			&= (t_2 - j + 1)/\log(n) \\
j  			&= t_2 - t\log(n) + 1 \\
dt 			&= -\frac{1}{\log(n)}dj \\
j  			&= 1 \Rightarrow t = \frac{t_2}{\log(n)} = \frac{\alpha \log(n)}{\log(n)} = \alpha \\
j  			&= t_2 \Rightarrow t = 1/\log(n) \rightarrow 0, \text{ as } n \rightarrow \infty \\
\gamma_{j}	&= \gamma_{t_2 - t\log(n) + 1} = \gamma_s + \frac{\gamma_f - \gamma_s}{t_2 - 1} (t_2 - t\log(n)) \\
			&\rightarrow \gamma_s + \frac{\alpha - t}{\alpha} (\gamma_f - \gamma_s), \text{ as } n \rightarrow \infty \\
\lambda_j 	&= \lambda(\gamma_j) \rightarrow \lambda \left( \gamma_s \frac{\alpha - t}{\alpha} (\gamma_f - \gamma_s) \right), \text{ as } n \rightarrow \infty
\end{aligned}
\right].
\end{equation*}

Let us denote $\gamma(t) = \gamma_s + \frac{\alpha - t}{\alpha} (\gamma_f - \gamma_s)$. We also want to evaluate $\log(\prod_{k = j}^{t_2}(\gamma_k))$. First of all we have

\begin{equation}
\label{eq:GammaProd}
\begin{split}
\prod_{k = j}^{t_2} \gamma_k &= d \cdot \frac{\gamma_j}{d} \cdot d \cdot \left( \frac{\gamma_j}{d} + 1 \right) \cdots d \cdot \left( \frac{\gamma_j}{d} + t_2 - j  \right)\\
&= d^{t_2 - j + 1} \frac{\Gamma \left( \frac{\gamma_j}{d} + t_2 - j + 1 \right)}{\Gamma \left( \frac{\gamma_j}{d} \right)}. 
\end{split}
\end{equation}

Let us denote $t' = t\log(n)$. Since $\gamma_j/d = (\gamma_s + (j - 1)d)/d = \gamma_s/d + j - 1$ we can rewrite \eqref{eq:GammaProd} as

\begin{equation}
\label{eq:GammaProdModified}
d^{t'} \frac{\Gamma \left( \frac{\gamma_s}{d} + t_2 \right) }{\Gamma \left( \frac{\gamma_s}{d} + t_2 - t' \right)} = d^{t'} \frac{\Gamma \left( \frac{\gamma_f}{d} \right) }{\Gamma \left( \frac{\gamma(t)}{d} \right)}.
\end{equation}

The natural logarithm of the gamma function is equal to

\begin{equation*}
\ln \left( \Gamma(z) \right) = z(\ln(z) - 1) + \mathcal{O} (\log(z)).
\end{equation*}

Thus, the dominant part of \eqref{eq:GammaProdModified} can be written as

\begin{equation}
\begin{split}
\log &\left( d^{t'} \frac{\Gamma \left( \frac{\gamma_f}{d} \right) }{\Gamma \left( \frac{\gamma(t)}{d} \right)} \right) \\
= &\bigg(  t'\ln(d) + \frac{\gamma_f}{d} \left( \ln \left( \frac{\gamma_f}{d} \right)  -1 \right) \\
&- \frac{\gamma(t)}{d} \left( \ln \left( \frac{\gamma(t)}{d} \right)  - 1 \right) \bigg)/\ln(2) \\
= &\left( \frac{\gamma_f \ln(\gamma_f) - \gamma(t) \ln(\gamma(t))}{d} - t' \right)/\ln(2) \\
= &\left( \frac{\gamma_f \ln(\gamma_f) - \gamma(t) \ln(\gamma(t))}{\gamma_f - \gamma_s} \frac{\alpha}{t} - 1 \right)t\log(n)/\ln(2).
\end{split}
\end{equation}

Now, the sum in \eqref{eq:SumProduct} approaches the following double integral as $n$ approaches infinity.

\begin{equation*}
\int_{0}^{\alpha} \frac{1}{\left(t \cdot \ell (t) + c_s\right)^2} \exp (I(t; \alpha, \gamma_s, \gamma_f)) dt,
\end{equation*}

where

\begin{equation*}
I(t; \alpha, \gamma_s, \gamma_f) = \int_{0}^{t} \frac{\log(\gamma(s)) - \lambda(\gamma(s))}{s\ell(s) + c_s} ds,
\end{equation*}

and

\begin{equation*}
\ell(s) = \left( \frac{\gamma_f \ln(\gamma_f) - \gamma(t) \ln(\gamma(t))}{\gamma_f - \gamma_s} \frac{\alpha}{s} - 1 \right)/\ln(2).
\end{equation*}

Now, let $i = t_2$ in \eqref{eq:stepJ} to get

\begin{equation}
\label{eq:Nexpression}
N = \sum_{j = 1}^{t_2} n_j = \frac{cn - (\log(\gamma_f) + c_s\log(n)) n_{t_2}}{\lambda_f}.
\end{equation}

The dominant part of this expression can be written as

\begin{equation*}
cn \int_{0}^{\alpha} \frac{c_s}{\left(t \cdot \ell (t) + c_s\right)^2} \exp (I(t; \alpha, \gamma_s, \gamma_f)) dt.
\end{equation*}

Like in previous derivations $t_1$ steps of plain-BKW with step-size $b$ is in total equal to

\begin{equation*}
t_1 \cdot b = (2(c_q - c_s) + 1 - \alpha) \frac{cn}{c_q}.
\end{equation*}

We have $n = N + t_1 \cdot b$. Using the expression for $N$ from \eqref{eq:Nexpression} and solving for $c$ finally gives us

\begin{equation*}
\resizebox{.98\hsize}{!}{
$\left( \frac{2(c_q - c_s) + 1 - \alpha}{c_q} + \int_{0}^{\alpha} \frac{c_s}{\left(t \cdot \ell (t) + c_s\right)^2} \exp (I(t; \alpha, \gamma_s, \gamma_f)) dt \right)^{-1}.$}
\end{equation*}

\end{IEEEproof}
	
}

\end{document}